# LEARN: Learned Experts' Assessment-based Reconstruction Network for Sparse-data CT

Hu Chen, Yi Zhang*, *Member, IEEE*, Yunjin Chen, Junfeng Zhang, Weihua Zhang, Huaiqiang Sun, Yang Lv, Peixi Liao, Jiliu Zhou, *Senior Member, IEEE*, and Ge Wang, *Fellow, IEEE*

*Abstract*—Compressive sensing (CS) has proved effective for tomographic reconstruction from sparsely collected data or under-sampled measurements, which are practically important for few-view CT, tomosynthesis, interior tomography, and so on. To perform sparse-data CT, the iterative reconstruction commonly uses regularizers in the CS framework. Currently, how to choose the parameters adaptively for regularization is a major open problem. In this paper, inspired by the idea of machine learning especially deep learning, we unfold a state-of-the-art "fields of experts" based iterative reconstruction scheme up to a number of iterations for data-driven training, construct a Learned Experts' Assessment-based Reconstruction Network (LEARN) for sparse-data CT, and demonstrate the feasibility and merits of our LEARN network. The experimental results with our proposed LEARN network produces a superior performance with the well-known Mayo Clinic Low-Dose Challenge Dataset relative to several state-of-the-art methods, in terms of artifact reduction, feature preservation, and computational speed. This is consistent to our insight that because all the regularization terms and parameters used in the iterative reconstruction are now learned from the training data, our LEARN network utilizes application-oriented knowledge more effectively and recovers underlying images more favorably than competing algorithms. Also, the number of layers in the LEARN network is only 50, reducing the computational complexity of typical iterative algorithms by orders of magnitude.

This work was supported in part by the National Natural Science Foundation of China under Grants 61671312, 31700858, 61302028 and in part by the National Institute of Biomedical Imaging and Bioengineering/National Institutes of Health under Grants R01 EB016977 and U01 EB017140. Asterisk indicates corresponding author.

H. Chen, Y. Zhang*, W. Zhang, and J. Zhou are with the College of Computer Science, Sichuan University, Chengdu 610065, China (e-mail: huchen@scu.edu.cn; yzhang@scu.edu.cn; zhangweihua@scu.edu.cn; zhoujl@scu.edu.cn).

Y. Chen is with ULSee Inc. Hangzhou 310020, China. (e-mail: chenyunjin_nudt@hotmail.com)

J. Zhang is with School of Computer and Information Engineering, Henan University of Economics and Law, Zhengzhou 450046, China. (e-mail: zhangjf20008@163.com)

H. Sun is with Department of Radiology, West China Hospital of Sichuan University, Chengdu 610041, China. (e-mail: sunhuaiqiang@scu.edu.cn)

Yang Lv is with Shanghai United Imaging Healthcare Co., Ltd, Shanghai 210807, China. (e-mail: yang.lv@united-imaging.com)

P. Liao is with Department of Scientific Research and Education, The Sixth People's Hospital of Chengdu, Chengdu 610065, China. (e-mail: universe6527@163.com)

G.Wang is with Department of Biomedical Engineering, Rensselaer Polytechnic Institute, Troy, NY 12180 USA (e-mail: wangg6@rpi.edu).

## I. INTRODUCTION

SPARSE-data CT is a fascinating topic that is both academically and clinically important. Academically speaking, tomographic image reconstruction from under-sampled data was previously considered infeasible, prohibited by the requirement imposed by the classic Nyquist sampling theorem. Thanks to the compressive sensing (CS) theory, nowadays many ill-posed inverse problems including sparse-data CT problems can be effectively solved using CS techniques [1]. In its nutshell, the success of CS-inspired image reconstruction is due to the utilization of prior knowledge especially the fact that there are major sparsity and correlation properties for many images including CT images. As a result, although an image volume has an apparent high dimensionality, it actually stays on a very low dimensional manifold, and a meaningful image reconstruction can be done on this manifold from much fewer data points. Clinically speaking, sparse-data CT solutions can enable a number of important diagnostic and interventional applications. Some futuristic cardiac CT architectures use a field-emission-based source ring, which represents a few-view CT configuration [2]. C-arm-based CT scans are valuable for surgical guidance and radiation treatment planning [3, 4]. Tomosynthesis and limited-angle tomography are also examples of sparse-data CT [5].

Due to the incompleteness and noise of projections for sparse-data CT, brute-force analytic algorithms produce severe image artifacts rendering resultant images useless, and iterative techniques are usually utilized to perform image reconstruction. For this purpose, many efforts were made over the past decades. Well-known algorithms include algebraic reconstruction technique (ART) [6], simultaneous algebraic reconstruction technique (SART) [7], expectation maximization (EM) [8], and so on. However, when the projection measurements are highly under-sampled, it is very difficult or impossible to achieve a satisfactory and stable solution with any prior information.

Compressive sensing (CS) is a breakthrough in solving under-determined inverse systems especially sparse-data CT [1]. Once the sampling process meets the restricted isometry property (RIP), there is a high possibility for CS to accurately reconstruct the original signal beyond the Nyquist sampling rate, aided by a proper sparsifying transform. The critical step is to find the sparsifying transform as the regularization term in the iterative reconstruction (IR) framework. Inspired by the theory of CS, Hu et al. and Sidky et al. used the discrete gradient transform, also termed as the total variation (TV) as the regularization term and obtained promising results [9-11]. However, as shown by Yu and Wang [12], the TV minimization is under the piecewise constant assumption that is generally unsatisfied in clinical practice, which means that TV compromises structural details and suffers from notorious blocky artifacts. To remedy this problem, many variants were proposed [13-16], and other substitutions were suggested to impose the sparsity prior into the iterative reconstruction framework, such as nonlocal means [17, 18], tight wavelet frames [19], dictionary learning [20-23], low rank [24], Gamma regularization (Gamma-Reg) [25, 26], and so on.

Most interestingly, via learning sparsifying transforms [27], Zheng et al. and Chun et al. combined penalized weighted-least squares and with sparsifying transform regularizations trained with high-quality external CT images for sparse-view and low-dose CT [28, 29]. Although CS-based iterative reconstruction methods achieved encouraging results, there are several drawbacks: (a) the iterative algorithms are time-consuming. The iterative procedure needs to repeat the projection and backprojection operations many times at a high computational cost. Meanwhile, calculating the gradients of the objective function including regularization terms further aggravate the burden; (b) for different clinical CT imaging tasks, it is very difficult to find a universal regularization term for consistently superior performance; and (c) there are multiple parameters to balance the data fidelity and regularization terms, and these parameters cannot be easily set. To address these drawbacks, in several recent studies the idea of learning from external datasets was introduced to mitigate the problem (b) to a certain degree, but other two issues (a) and (c) have not been attempted, to our best knowledge.

Recently, deep learning (DL) has drawn an overwhelming attention [30]. Until now, machine learning especially deep learning were mainly utilized for medical image analysis, such as organ segmentation [31], nodule detection [32], nuclei classification [33]. However, inspired by the fruitful results gained in the realm of low-level image processing [34-37], like image denoising, inpainting, deblurring or super-resolution, major efforts are being made in our field to reconstruct tomographic images using deep learning techniques [38-44]. Particularly, in [38] Chen et al. proposed a three-layers convolutional neural network (CNN) for noise reduction in low-dose CT. Kang et al. transformed low-dose CT (LDCT) images into the wavelet domain for deep learning based denoising [39]. Yang et al. observed that deep CNN with pixel-wise mean squared error (MSE) overly smoothened images, and proposed a perceptual similarity measure to measure the loss [40]. Inspired by the idea behind the autoencoder, Chen et al. developed a residual encoder decoder CNN (RED-CNN) for low-dose CT (LDCT) image denoising [41]. To suppress the artifacts from under-sampling for CT imaging, Han [42] and Jin et al. [43] independently proposed two U-Net based algorithms. More recently, generative adversarial networks (GANs) were introduced for low-dose CT [44]. In the Fully3D 2017 Conference, tomographic image reconstruction via deep learning (DL) is a highlight. Du et al. embedded a sparse prior trained through a K-sparse autoencoder (KSAE) into a classic iterative reconstruction framework [45, 46]. Researchers from KAIST utilized a U-Net architecture to deal with different topics in the image domain; e.g., sparse-view, limited-angle and low-dose CT [47-49]. Based on a denoising autoencoder, Li and Mueller proposed a symmetric network incorporating a residual block to suppress artifacts due to sparse sampling [50]. Another project presented by Cheng et al. from GE's group used DL to accelerate the convergence of iterative reconstruction [51], similar to the idea mentioned in [52]. In their study, intermediate (after 2 or 20 iterations) and corresponding final results (after 200 iterations) were used to train a neural network which significantly accelerated the reconstruction process.

Although these initial results with deep learning techniques are encouraging, they are the post-processing methods, and inherently overlook the data consistence. Here we see an opportunity to combine the deep learning techniques and iterative reconstruction algorithm for improved image reconstruction from sparse data. Until now, very limited results were presented in this aspect. Wang et al. proposed an accelerating MRI reconstruction strategy by imposing a deep learning based regularization term [53]. Based on the work of sparse coding [54], Yang et al. unfolded the alternating direction method of multipliers (ADMM) into a CNN network, efficiently accelerating the MRI reconstruction [55]. Similar to this work, a variational model was embedded into an unrolled gradient decent scheme for CS-based MRI reconstruction [56]. In [57], undersampled k-space data were utilized, and the zero-padded parts were replaced with predicted fully-sampled data from a trained image-to-image network. In [58], a CS based MRI reconstruction method was adapted into the GAN framework. To ensure the learned manifold is data consistent, a least-square penalty was introduced into the training process. In the field of CT reconstruction directly from sinogram data, to our best knowledge, the only work with deep learning is to expand FBP into a three-layer network, which learns the weightings and additional filters to reduce the error of a limited angle reconstruction [59]. However, when the sampling rate is low, the FBP network would fail to yield usable images.

Extending the prior results [54-56], in this paper we generalize the iteration reconstruction framework into a Learned Experts' Assessment-based Reconstruction Network (LEARN). There are three major benefits from our efforts:

a) The reconstruction procedure is fully neural-networked and significantly accelerated. The iterative procedure is first unfolded into a recurrent residual network. By fixing an appropriate number of iterations, the network is casted into a CNN-based network. By feeding this network with projection data directly, we only perform a limited number of forward computational steps instead of hundreds of iterations.

b) All the regularization terms and balancing parameters can be adaptively learned in the training stage. In the unfolded network, the regularization terms and balancing parameters become iteration-dependent parameters of a neural network, which means that these parameters can vary with each iteration.

This characteristic makes our model more flexible and more robust than other types of iterative reconstruction methods.

c) As shown below in detail, the image quality of the LEARN network is superior to or competitive with the state-of-the-art iterative methods at a much-reduced computational cost.

We also note that our LEARN is different from the recently proposed KSAE [46]. KSAE follows the classical IR steps with a data fidelity term and a regularization term. The main innovation of KSAE is to include a sparse prior in the form of a learned image-to-image mapping. Different from the LEARN network, the KSAE network was trained by LDCT images and their corresponding NDCT images, and the learned mapping function is fixed for the entire iterative procedure. [57] and [58] also trained the networks but only with image samples. There are some differences between these two studies: [57] kept the nonzero-padded parts unchanged, and [58] introduced a residual block. The proposed LEARN is directly trained by projection data and the corresponding NDCT images, and the learned regularization terms and balancing parameters are specific to each iteration.

The rest of the paper is organized as follows. In the next section, we derive and explain our proposed LEARN network. In the third section, we describe the experimental design and analyze representative results. In the last section, we discuss relevant issues and conclude the paper.

## II. METHODS

### A. Regularized CT Reconstruction

Typically, the CT reconstruction problem is treated as solving a linear system:

$$Ax = y \quad (1)$$

where $x = (x_1, x_2, ..., x_J)^T$ denotes a vector of discrete attenuation coefficients for a patient image, $A$ is the imaging system or projection matrix of $I \times J$ elements corresponding to a specific configuration of the CT system, and $y = (y_1, y_2, ..., y_I)^T$ represents the measured data after calibration and log-transform. Mathematically, the element of $A$, $a_{i,j}$, stands for the intersection of the $i$-th x-ray path with the $j$-th pixel. The purpose of image reconstruction is to recover the unknown $x$ from the system matrix $A$ and observed data $y$.

If a set of projection data is complete without significant noise, (1) can be analytically inverted with FBP in either fan-beam or cone-beam geometry [60]. However, for the sparse-data CT reconstruction problem, the linear system (1) becomes underdetermined and has infinite solutions. The reconstructed image with FBP will suffer from strong image artifacts and significantly degraded image quality, and iterative reconstruction algorithms are the method of choice to overcome the challenges because these algorithms can easily accommodate imaging physics and prior knowledge at the cost of much-increased computational time.

For iterative image reconstruction, (1) can be solved by minimizing the following constrained objective function:

$$x = \arg\min_x E(x) = \arg\min_x \|Ax - y\|_2^2, \quad s.t. \ x_j \geq 0 \quad \forall j, \quad (2)$$

where $\|\cdot\|_2^2$ denotes the $L_2$ norm. Popular iterative methods, such as ART, SART and EM, can be employed to solve (2), but artifacts may still exist when (2) is not well posed. To address this problem, various prior knowledge can be incorporated into (2) for regularization. Then, a regularized objective function is expressed as

$$\begin{aligned} x &= \arg\min_x E(x) \\ &= \arg\min_x \frac{\lambda}{2}\|Ax - y\|_2^2 + R(x), \quad s.t. \ x_j \geq 0 \quad \forall j, \end{aligned} \quad (3)$$

where the first term is for data fidelity, which addresses the consistency between reconstructed $x$ and observed measurement $y$, the second term $R(x)$ is for regularization, and $\lambda$ controls the balance between data fidelity and regularization.

Previous studies were mainly focused on the development and implementation of different prior terms. For example, TV is a popular one for its ability of keeping sharp discontinuities but it is actually based on the piecewise constant assumption for an underlying image [61]. In biomedicine, it is generally inaccurate to assume that CT images are piecewise constant. Better alternative regularizers include various variants of TV and other regularizers, such as total generalized variation (TGV) [16], nonlocal TV (NLTV) [62], and tight framelet (TF) [63], but most of them were handcrafted and cannot be used for all kinds of images in different clinical applications.

### B. LEARN Network for Sparse-Data CT

In [64], a generalized regularization term, referred to as fields of experts (FoE), was proposed as

$$R(x) = \sum_{k=1}^{K} \phi_k(G_k x), \quad (4)$$

where $K$ is the number of regularizers, $G_k$ is a transform matrix of size $N_f$, which can be seen as a convolutional operator for a CT image $x$, and $\phi_k(\cdot)$ is a potential function. In the FoE model, both $G_k$ and $\phi_k(\cdot)$ can be learned from training data. Inserting (4) into (3), we have

$$\begin{aligned} x &= \arg\min_x E(x) \\ &= \arg\min_x \frac{\lambda}{2}\|Ax - y\|_2^2 + \sum_{k=1}^{K} \phi_k(G_k x), \quad s.t. \ x \geq 0. \end{aligned} \quad (5)$$

where $\lambda$ is the weighting parameter for the data fidelity term.

Assume that the second term in (5) is differentiable and convex. Then, a simple gradient descent scheme can be applied to optimize (5):

$$x^{t+1} = x^t - \alpha \cdot \eta(x^t) = x^t - \alpha \frac{\partial E}{\partial x}, \quad (6)$$

where $\alpha$ is the search step and should be carefully chosen. With (5), the gradient term can be obtained as

$$\eta(x^t) = \lambda A^T(Ax^t - y) + \sum_{k=1}^{K} (G_k)^T \gamma_k(G_k x^t), \quad (7)$$

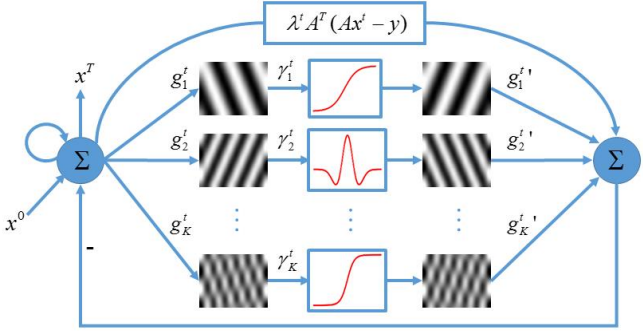

Fig. 1. Network architecture corresponding to (8).

where $\gamma(\cdot)=\phi'(\cdot)$, the superscript $T$ stands for the transpose of a matrix, and $A^T$ denotes the back projection operator. The transpose of the filter kernel can be obtained after the convolutional kernel is rotated by 180 degrees. In our previous work [65] for low-level vision tasks, a time-dependent version was proposed. The balancing parameters, potential functions and kernels can be changed iteration-wise, which makes our method more flexible. This method has achieved promising results in several important tasks, including image denoising, inpainting, and MRI reconstruction. To our best knowledge, our work is the first attempt to apply this strategy to CT reconstruction.

By letting the terms of (7) be iteration-dependent, (6) is changed to

$$x^{t+1} = x^t - \left( \lambda^t A^T (Ax^t - y) + \sum_{k=1}^{K} (G_k^t)^T \gamma_k^t (G_k^t x^t) \right). \quad (8)$$

Since $\eta(x^t)$ can be freely scaled, $\alpha$ is neglected in (8). All the symbols with the superscript $t$ signifies an iteration-dependency. If $\lambda^t$, $G_k^t$ and $\gamma_k^t$ keep fixed in (8), the iteration returns to the original FoE prior. Meanwhile, TV, wavelet, and other hand-crafted regularization terms can be seen as special cases of the FoE priors.

Our method has generalized these CS-based iterative reconstruction models. The critical part to solve (8) is to determine the specific forms of $G_k^t$ and $\gamma_k^t$. In the reference [64], the FoE prior was learned from training data by hybrid Monte Carlo sampling, which was based on an idea similar to learning a sparse transform for CS-based reconstruction [27-29]. Although learned prior could improve the performance to a certain extent, the iterative procedure is still time-consuming.

Actually, the term $\sum_{k=1}^{K} (G_k^t)^T \gamma_k^t (G_k^t x^t)$ in (8), with the transforms dependent on the iteration index, can be interpreted as being parallel to a classical CNN. In each iteration, an image $x$ is convolved with a set of linear filters, which can be treated as a recurrent residual CNN. There are two convolution layer, $(G_k^t)^T$ and $G_k^t$, and one activation function layer $\gamma_k^t$. Meanwhile, the image $x^{t-1}$ in the previous iteration is involved to update the new one $x^t$, which works in the principle similar to that of the residual network. Based on the above observation, the transform matrices $G_k^t$ and $(G_k^t)^T$ can be substituted by the corresponding convolutional kernels $g_k^t$ and $g_k^t{'}$ respectively. Hence, (8) can be represented as a CNN shown in Fig. 1, where the loop denotes the iterative procedure.

To extend the flexibility of the network, we actually implement the term $\sum_{k=1}^{K} (G_k^t)^T \gamma_k^t (G_k^t x^t)$ in each iteration as a multilayer CNN, including convolutional and ReLU layers. With a predetermined number of iterations, we can unfold Fig. 1 to a deep CNN, with all the regularization terms and parameters trainable. In this sense, our proposed network to solve (8) is named as a Learned Experts' Assessment-based Reconstruction Network ("LEARN"), as illustrated in Fig. 2.

In Fig. 2, we use a three layer CNN (3Layer-CNN) [38] to substitute $\sum_{k=1}^{K} (G_k^t)^T \gamma_k^t (G_k^t x^t)$ for each iteration. 3Layer-CNN was proposed to denoise low-dose CT images, which is a post-processing method. The mapping function of 3Layer-CNN can be formulated by

$$M(x^{t-1}) = \mathbf{W}_3^{t-1} * \text{ReLU}(\mathbf{W}_2^{t-1} * \\ \text{ReLU}(\mathbf{W}_1^{t-1} * x^{t-1} + \mathbf{b}_1^{t-1}) + \mathbf{b}_2^{t-1}) + \mathbf{b}_3^{t-1}, \quad (9)$$

where the weights $\mathbf{W}_1^{t-1}$, $\mathbf{W}_2^{t-1}$, $\mathbf{W}_3^{t-1}$ consist of $n_1$, $n_2$ and $n_3$ convolution kernels with a uniform size of $s_1$, $s_2$ and $s_3$ respectively, $\mathbf{b}_1^{t-1}$, $\mathbf{b}_2^{t-1}$ and $\mathbf{b}_3^{t-1}$ are the corresponding biases, $*$ represents the convolution operator, and ReLU$(\cdot)$ is the activation function. The whole LEARN network is cascaded with multiple iterations denoted by blocks in Fig. 2. The zoomed green box shows the flowchart of operations in each block. In addition to the 3Layer-CNN at the bottom, a term $\lambda^t A^T (Ax^t - y)$ corresponding to the data fidelity term, and a shortcut connection from $x^{t-1}$ to $x^t$. All these links are summed into the intermediate reconstruction $x^t$. While the stacked 3Layer-CNNs alone are subject to the risk of over-smoothing the image details and difficulty of training, the overall architecture is a residual network that preserves the structural details and accelerates the training speed [41].

It is underlined that all the iteration-index-dependent parameters of the LEARN network, including the convolution operators, will be learned from training data. The numbers of filters $\{n_1, n_2, n_3\}$, the kernel sizes $\{s_1, s_2, s_3\}$, and the total number of iterations $N_t$ are manually set in this pilot study. The initial inputs to the network include $x^0$, $A$, $A^T$ and $y$, and the corresponding final output is the reconstructed image $x^{N_t}$. The input to the network $x^0$ can be set to 0 or an approximate reconstruction such as that obtained with FBP or a popular IR method. The data fidelity term is utilized in every iteration. The convolution operations are performed on intermediate results in the image domain.

### C. Training the LEARN Network

The proposed LEARN network can be trained in a supervised manner, which means a training dataset should be prepared with both undersampled measurements and paired high-quality CT images. Specifically, the training dataset $D$ consists of $N_D$ samples $(y_s, x_s)_{s=1}^{N_D}$, where $y_s$ is undersampled measurements, and $x_s$ is the corresponding reference image. The parameter set $\Theta^t = \{\lambda^t, \mathbf{W}_1^t, \mathbf{W}_2^t, \mathbf{W}_3^t, \mathbf{b}_1^t, \mathbf{b}_2^t, \mathbf{b}_3^t\}$ contains all the parameters that are iteration-index-specific, including 1) the balancing parameter $\lambda^t$, 2) the filter weights $\{\mathbf{W}_1^t, \mathbf{W}_2^t, \mathbf{W}_3^t,\}$ and biases $\{\mathbf{b}_1^t, \mathbf{b}_2^t, \mathbf{b}_3^t\}$. The training model is formulated to minimize the loss function in the form of accumulated mean squared errors (MSE):

$$\begin{cases} \min_{\Theta} L(D;\Theta) = \dfrac{1}{2N_D} \sum_{s=1}^{N_D} \left\| x_s^{N_t}(y_s, \Theta) - x_s \right\|_2^2, \\ s.t. \begin{cases} x_s^{t+1} = x_s^t - \left(\lambda^t A^T (A x_s^t - y_s) + M(x_s^t)\right), \\ t = 0, 1, 2 \ldots, N_t - 1 \end{cases} \end{cases} \quad (10)$$

where $x_s^{N_t}(y_s, \Theta)$ denotes the reconstructed image after the final iteration $N_t$ from under-sampled data $y_s$. In this study, we optimized the loss function using the Adam method [66]. For the initialization of training parameters, the base learning rate was set to $10^{-4}$, and slowly decreased down to $10^{-5}$. The convolution kernels were initialized according to the random Gaussian distribution with zero mean and standard deviation 0.01. The initialization of the network parameters may influence the training and the performance of LEARN. Although the optimization of the training parameter initialization was not the main point in this paper, some analyses were given in the supplemental materials.

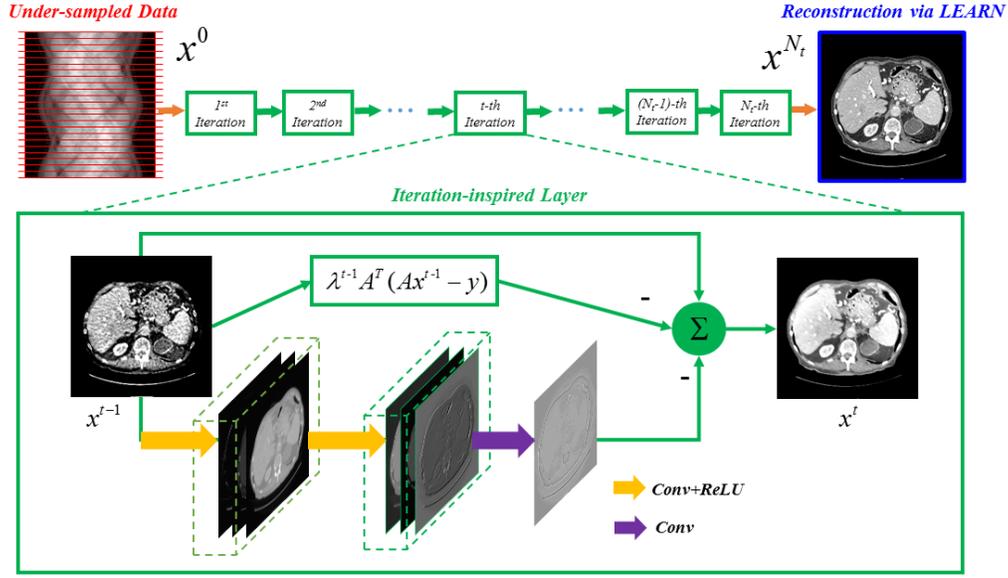

Fig. 2. Overall structure of our proposed LEARN network.

To perform the back propagation procedure for the proposed LEARN network, the gradient computation is the key but it is different from that for a normal CNN due to the existence of the data fidelity term. The formulation of all the computational steps for back propagation involves the applications of the chain rule multiple times for every layer but they are technically trivial. For brevity, here we only choose the differentiation parts as an example to illustrate how to perform back propagation, and the other CNN layers can be similarly computed.

The associated gradient $\partial L / \partial \Theta^t$ can be obtained by back propagation as follows:

$$\frac{\partial L}{\partial \Theta^t} = \frac{\partial x^{t+1}}{\partial \Theta^t} \cdot \frac{\partial x^{t+2}}{\partial x^{t+1}} \cdots \frac{\partial x^{N_t}}{\partial x^{N_t-1}} \frac{\partial L}{\partial x^{N_t}}. \quad (11)$$

Clearly, we need to compute the three kinds of differentiation in (11) respectively; i.e., $\partial x^{t+1}/\partial \Theta^t$, $\partial x^{t+2}/\partial x^{t+1}$ and $\partial L/\partial x^{N_t}$. First, according to (10) we compute the gradient of the loss function $L$ with respect to the reconstructed result $x^{N_t}$ as

$$\frac{\partial L}{\partial x^{N_t}} = x^{N_t} - x. \quad (12)$$

Second, due to $\Theta^t = \{\lambda^t, \mathbf{W}_1^t, \mathbf{W}_2^t, \mathbf{W}_3^t, \mathbf{b}_1^t, \mathbf{b}_2^t, \mathbf{b}_3^t\}$, $\partial x^{t+1}/\partial \Theta^t$ can be respectively computed according to (10). For brevity, only $\partial x^{t+1}/\partial \lambda^t$, $\partial x^{t+1}/\partial \mathbf{W}_1^t$ and $\partial x^{t+1}/\partial \mathbf{b}_1^t$ are given here (other derivatives can be similarly obtained). The derivative of $x^{t+1}$ with respect to $\lambda^t$ is expressed as

$$\frac{\partial x^{t+1}}{\partial \lambda^t} = -\left(A^T(A x_s^t - y_s)\right)^T, \quad (13)$$

the derivative of $x^{t+1}$ with respect to $\mathbf{W}_1^t$ is expressed as

$$\frac{\partial x^{t+1}}{\partial \mathbf{W}_1^t} = -\frac{\partial M(x^t)}{\partial \mathbf{W}_1^t}, \quad (14)$$

and the derivative of $x^{t+1}$ with respect to $\mathbf{b}_1^t$ is written as

$$\frac{\partial x^{t+1}}{\partial \mathbf{b}_1^t} = -\frac{\partial M(x^t)}{\partial \mathbf{b}_1^t}. \quad (15)$$

Third, according to the formula for updating $x^{t+1}$ in (10), $\partial x^{t+2}/\partial x^{t+1}$ is given as

$$\frac{\partial x^{t+2}}{\partial x^{t+1}} = I - (\lambda^{t+1} A^T A + \frac{\partial M(x^{t+1})}{\partial x^{t+1}}), \quad (16)$$

where $I$ is an identity matrix of $\sqrt{J} \times \sqrt{J}$ and $\partial M(x^{t+1})/\partial x^{t+1}$, $\partial M(x^{t+1})/\partial \mathbf{W}_1^t$ and $\partial M(x^{t+1})/\partial \mathbf{b}_1^t$ are the derivative of the CNN-based mapping function in (9), which can be calculated in the standard back propagation procedure for a classical CNN. With (12) to (16), all the terms in (11) can be calculated. As a result, $\partial L/\partial \lambda^t$, $\partial L/\partial \mathbf{W}_1^t$ and $\partial L/\partial \mathbf{b}_1^t$ can be respectively obtained.

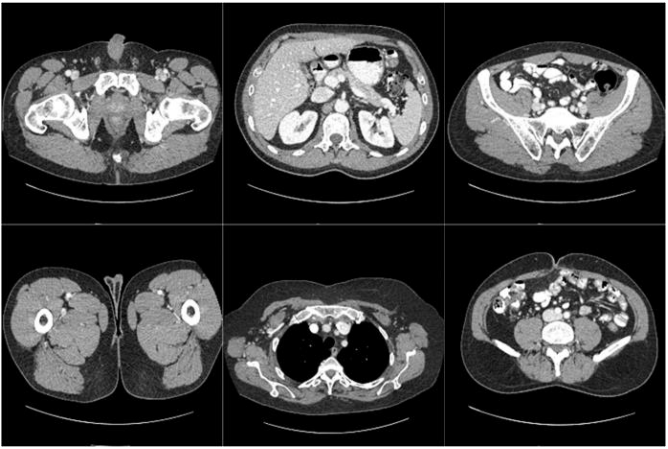

Fig. 3. Examples in the dataset. The display window is [-150 250] HU.

III. EXPERIMENTAL DESIGN AND REPRESENTATIVE RESULTS

To evaluate the imaging performance of the LEARN network under realistic conditions, a set of clinical data and images was used, which was established and authorized by Mayo Clinics for "*the 2016 NIH-AAPM-Mayo Clinic Low Dose CT Grand Challenge*". The image dataset contains 5,936 1mm thickness full dose CT images from 10 patients. Refer to [67] for more details about this dataset. The projection dataset is composed of projection data from 2,304 views per scan. The reference images were generated using the FBP method from all 2,304 projection views. The projection data was down-sampled to 64 and 128 views respectively to simulate the few-view geometry. 25 images were randomly selected for each patient, and there are totally 250 images cases into our dataset for this study. Fig. 3 demonstrates examples in the dataset. It is seen that different parts of the human torso were included. Another observation is that due to the thin slice thickness, image noise is evident even in the fully sampled images. The LEARN network was trained with a subset of paired full dose and under-sampled images, totally 200 image pairs from 8 patients. The rest of the 50 image pairs from the other 2 patients were respectively used for testing. After the first random selection, the training set included 37 thoracic, 82 abdominal and 81 pelvic images. Then, the testing dataset included 8 thoracic, 18 abdominal and 24 pelvic images. For fairness, cross-validation was performed with the testing dataset.

In our experiments, the following basic parameters were evaluated for their impacts on image quality. The number of filters in the last layer $n_3$ was set to 1 and the numbers of filters in the first two layers were both set to 48. The kernel size of all layers was set to 5×5. All $\lambda^t$ in our network were initialized to 0 and the initial input to the network $x^0$ was set to the FBP result. The number of iterations $N_t$ was set to 50. The proposed LEARN network was implemented in MATLAB using MatConvNet [68] and all the experiments were performed in MatLab 2017a on a PC (Intel i7 6800K CPU and 64 GB RAM). The training stage is time-consuming on CPU. A common way for acceleration is to work in parallel on GPU. In our work, the training process was executed with a graphic processing unit card GTX Titan Xp. Our codes for this work are available on https://github.com/maybe198376/LEARN.

Three classic metrics, including the root mean square error (RMSE), peak signal to noise ratio (PSNR) and structural similarity index measure (SSIM) [69], were chosen for quantitative assessment of image quality.

Five state-of-the-art methods were compared against our LEARN network, including ASD-POCS [11], dual dictionary learning (Dual-DL) [20], total generalized variation based penalized weighted least-squares (PWLS-TGV) [16], Gamma-Reg [25, 26] and FBPConvNet [43]. ASD-POCS is a widely used iterative reconstruction method with the TV regularization. Dual-DL is a contemporary iterative reconstruction model aided by learned dictionaries from external data, which can be grouped into the category of learning-based methods. PWLS-TGV is a statistical IR method with a regularization term constructed with higher order derivatives. Gamma-Reg is a recently proposed IR method, which utilizes the Gamma regularization to simulate the fractional norm between the $l_0$-norm and $l_1$-norm. FBPConvNet is the most recently proposed CNN-based sparse-view CT method. It is essentially a post-processing method. The parameters of ASD-POCS, PWLS-TGV, Gamma-Reg and Dual-DL were optimized using a golden-section search to minimize RMSE. For fair comparison, the external global dictionary for Dual-DL was trained with the same training dataset as that used by the LEARN network. FBPConvNet was trained with the same training strategy in the original reference [43]. Totally, 500 images were selected, including the 200 images in the LEARN's training set. Meanwhile, data augmentation was applied, making the total number of training samples reach 2000, which is consistent with [43].

*A. Visualization-Based Evaluation*

To visualize the performance of our LEARN network, representative slices using the first random training and testing dataset were selected. In Fig. 4, the abdominal image reconstructed from 64 views were reconstructed using different methods. As the sampling rate was rather sparse, the artifacts in the resultant FBP reconstruction is too severe to show any diagnostically useful information. All the other four methods efficiently suppressed the artifacts. However, as shown in Fig. 4(c), ASD-POCS suffered from the notorious blocky effect, caused by the clinically improper assumption that the underlying image was piecewise constant [61]. Dual-DL produced a better visual effect than ASD-POCS, but the edges of tissues were blurred. The reason for blurring is the use of the weighted average of dictionary atoms. This procedure can efficiently remove the noise in smooth regions, but the details may not be kept very well. In Fig. 4(e) and (f), PWLS-TGV and Gamma-Reg mitigated the blocky effect to a certain extent, but the details in the liver were still noisy. It is observed that Fig. 4(g) has the best spatial resolution. The structures were clear, and even the noise in the reference image was eliminated. However, comparing to the reference image, many important details were smoothened away. Except enhanced blood vessels, some other structures were distorted as indicated by the red arrows. This phenomenon was also observed in [38, 41] and a brief analysis was given in our previous paper [41]. Three comments can be made on this defect. First, FBPConvNet involves multiple down-sampling and up-sampling operations. These operations may help enlarge the effective receptive field to extract global features of artifacts, but the images details may be missed during these operations. Second, learning-based post processing methods are heavily dependent on the training samples, and 500 samples may be a small number relative to the current capacity of FBPConvNet. Third, FBPConvNet only uses projection data to generate the pseudo-inverse, which led to the result close to that reconstructed via FBP. In other words, their work did not involve the measurements into the processing procedure, which means that the image quality could only be suboptimal. In Fig. 4(h), the LEARN network maintained most of the details, especially in terms of noise reduction, the contrast enhanced blood vessels and other small structures.

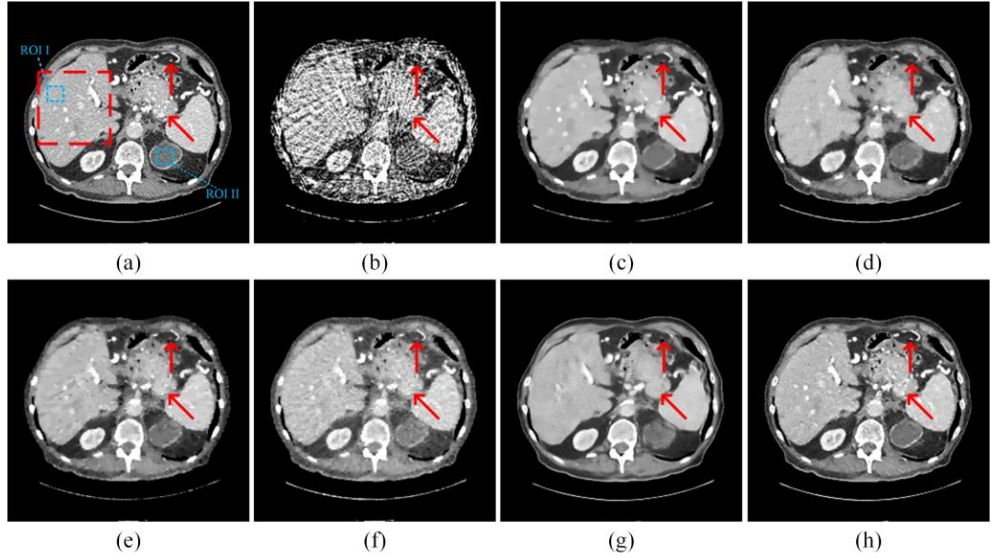

Fig. 4. Representative abdominal images reconstructed using various methods. (a) The reference image versus the images reconstructed using (b) FBP, (c) ASD-POCS ($\lambda = 0.07$), (d) Dual-DL ($n = 16$, $d_s = 2$, $\beta = 0.03$), (e) PWLS-TGV ($\beta_1 = 1 \times 10^{-2}$, $\beta_2 = 1.3 \times 10^{-4}$), (f) Gamma-Reg ($\lambda = 3 \times 10^{-3}$, $\alpha = 1.5$, $\beta = 8$), (g) FBPConvNet, and (h) LEARN respectively. The red arrows point to some key details, which can only be discriminated with the LEARN network. The red box labels a region of interest (ROI), which is magnified in Fig. 5. The display window is [-150 250] HU.

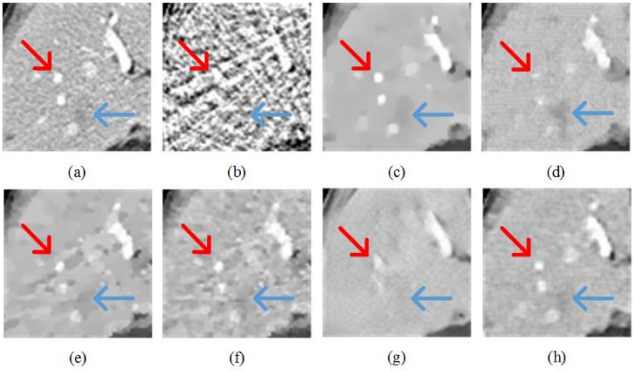

Fig. 5. Zoomed region of interest (ROI) marked by the red box in Fig. 4(a). (a) The reference image versus the images reconstructed using (b) FBP, (c) ASD-POCS, (d) Dual-DL, (e) PWLS-TGV, (f) Gamma-Reg, (g) FBPConvNet, and (h) LEARN respectively ((a)-(g) from Fig. 4(a)-(g)). The arrows indicate two locations with significant visual differences. The display window is [-150 250] HU.

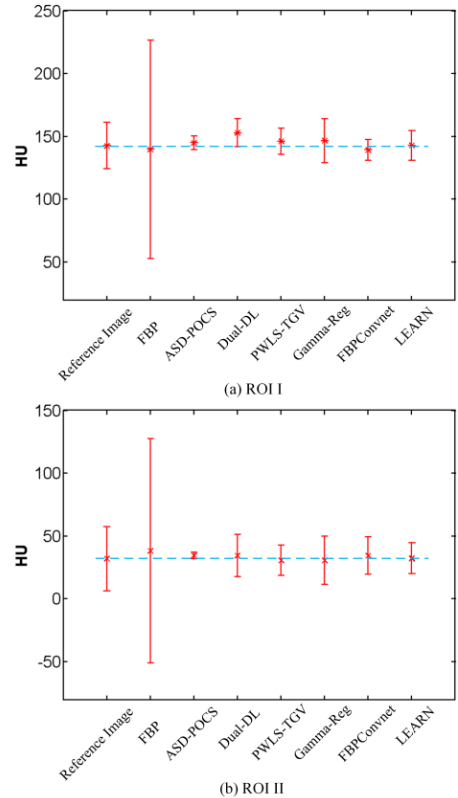

Fig. 6. Means and standard variations for (a) ROI I and (b) ROI II reconstructed using different methods.

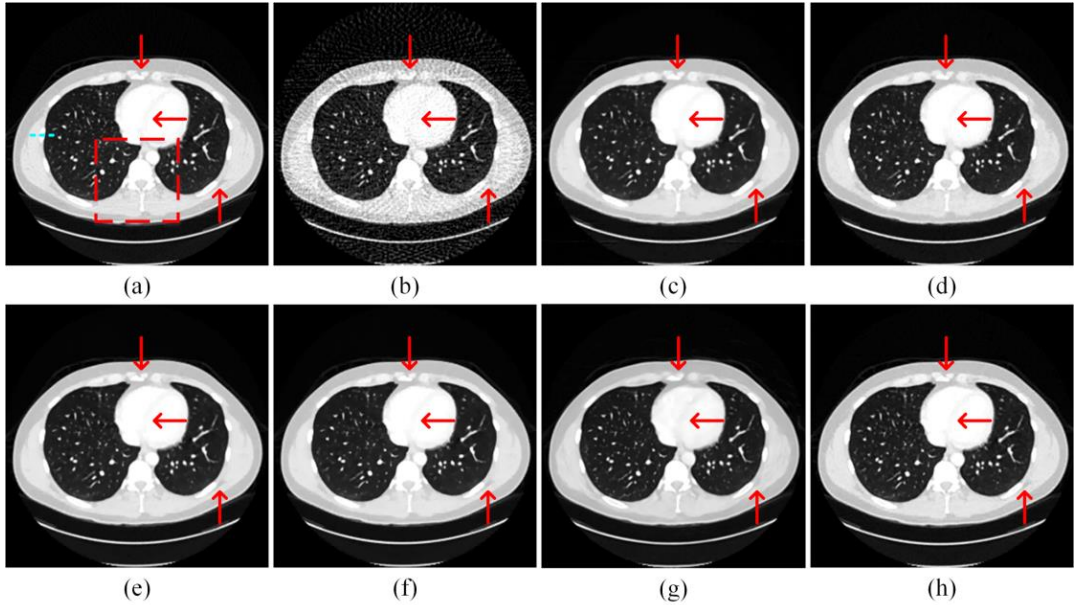

Fig. 7. Representative thoracic images reconstructed using various methods. (a) The reference image versus the images reconstructed using (b) FBP, (c) ASD-POCS ($\lambda = 0.05$), (d) Dual-DL ($n = 25$, $d_s = 2$, $\beta = 0.01$), (e) PWLS-TGV ($\beta_1 = 1 \times 10^{-2}$, $\beta_2 = 0.5 \times 10^{-4}$), (f) Gamma-Reg ($\lambda = 2 \times 10^{-3}$, $\alpha = 1.2$, $\beta = 7$), (g) FBPConvNet, and (h) LEARN respectively. The red arrows point to some details, which can be discriminated by the LEARN network. The red box labels an ROI to be magnified in Fig. 8. The profiles along the dotted blue line are in Fig. 9. The display window is [-1000 200] HU.

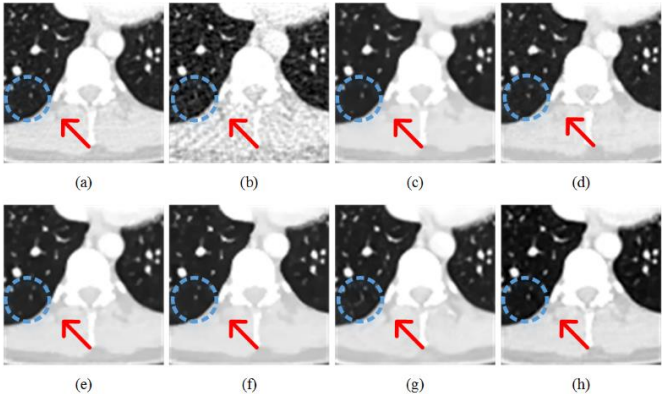

Fig. 8. Zoomed parts over the region of interest (ROI) marked by the red box in Fig. 6(a). (a) The reference image versus the images reconstructed using (b) FBP, (c) ASD-POCS, (d) Dual-DL, (e) PWLS-TGV, (f) Gamma-Reg, (g) FBPConvNet, and (h) LEARN respectively ((a)-(h) from Fig. 6(a)-(f)). The red arrows indicate a selected region for visual difference. The blue dotted circle shows another region, where the results of FBP and FBPConvNet generated similar artifacts. The display window is [-1000 200] HU.

Fig. 5 demonstrated the results in the zoomed ROI, which was indicated by the red box in Fig. 4(a). The red arrows indicated a liver region containing several contrast enhanced blood vessels. The blue arrow pointed to a location with possible metastasis, which is clinically important. In Fig. 5(b), the artifacts were severe and covered all the information. Although ASD-POCS preserved some structures, the details were heavily blurred. Dual-DL kept parts of the details, but the contrast was low with artifacts due to the weighted average of dictionary patches. PWLS-TGV did not remedy the blocky effect very well. The noise in Fig. 5(f) is still noticeable. FBPConvNet gave the best contrast in all the cases, but many details were smoothened in the liver. The possible metastasis was difficult to be recognized. In Fig. 5(h), the LEARN network preserved the vessels better than the other methods, with the metastasis being clearly identified. To validate whether the proposed method introduced HU bias, Fig. 6 shows the means and standard deviations in two (liver and kidney) homogeneous regions reconstructed using different methods, as indicated with the two blue boxes (ROI I and ROI II) in Fig. 4(a). It can be seen that LEARN had the mean closest to the reference image in both ROIs and its standard deviations were more consistent to that in the reference image. ASD-POCS, Dual-DL, PWLS-TGV and FBPConvNet had much smaller standard deviations than that in the reference image in ROI I, which can be viewed as an evidence of over-smoothing.

Fig. 7 presents the thoracic images reconstructed from 128 views using the different methods respectively. With the increase of sampling angles, the artifacts in the FBP reconstruction were significantly reduced than the counterpart in Fig. 4(b). All the other methods eliminated most of the artifacts. The red arrows indicated three regions with structural details. The Dual-DL, PWLS-TGV, Gamma-Reg and LEARN network reproduced images most consistent to the reference. ASD-POCS blurred the edges in the top of the image. FBPConvNet distorted the details in the top region in the image, and overly smoothened the interventricular septum in the middle of the image. We also chose a small region in Fig. 7 to enlarge more details for further examination in Fig. 8. As marked by the red arrows, the LEARN network preserved the edges better than the other methods. The blue circle indicates a pseudo-structure observed by FBP and FBPConvNet in Fig. 8(b) and (g) respectively. Actually, the FBP result was the input of FBPConvNet so that it can be predicted that without referencing to the original projection data, CNN-based post-processing methods cannot reliably distinguish between subtle details and weak artifacts. To further demonstrate the ability of our method for structure preservation, the horizontal profile was plotted as a dotted blue line in Fig. 9. It can be easily observed that the profile of LEARN is most consistent to the reference image across edges and in approximately homogeneous regions.

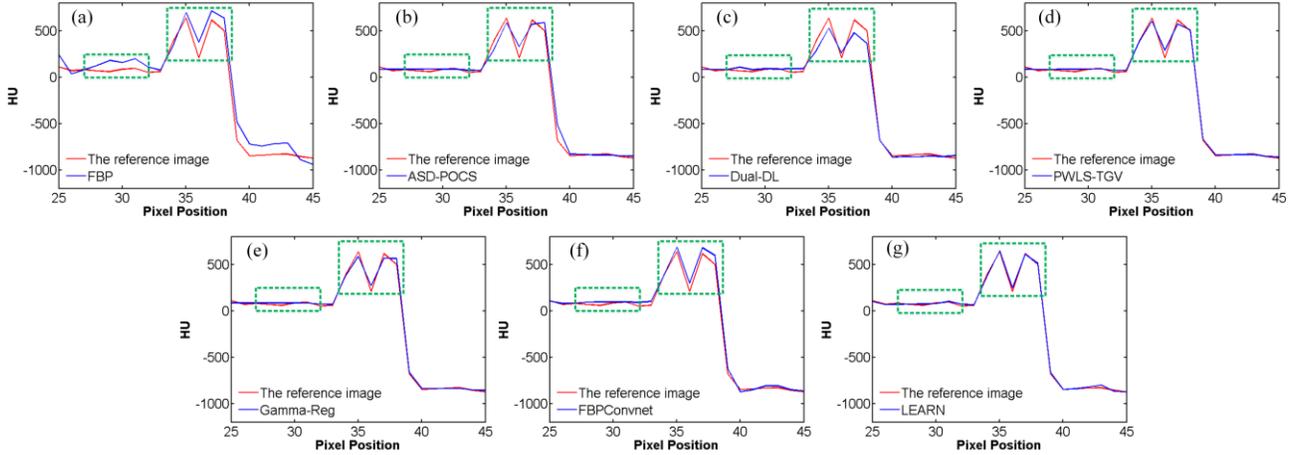

Fig. 9. The horizontal profiles along the dotted blue line in Fig. 7(a) of the reference image versus the images reconstructed using (a) FBP, (b) ASD-POCS, (c) Dual-DL, (d) PWLS-TGV, (e) Gamma-Reg, (f) FBPConvNet, and (g) LEARN respectively. Two dotted green boxes label approximately homogenous and edge-rich regions respectively.

Table I. QUANTITATIVE RESULTS ASSOCIATED WITH DIFFERENT ALGORITHMS IN THE ABDOMINAL CASE.

| No. of views | 64 | | | 128 | | |
|---|---|---|---|---|---|---|
| | RMSE | PSNR | SSIM | RMSE | PSNR | SSIM |
| FBP | 0.0546 | 25.2601 | 0.5815 | 0.0310 | 30.1841 | 0.7523 |
| ASD-POCS | 0.0223 | 32.0145 | 0.8542 | 0.0176 | 35.0914 | 0.9179 |
| Dual-DL | 0.0230 | 32.7584 | 0.8670 | 0.0158 | 35.5487 | 0.9225 |
| PWLS-TGV | 0.0210 | 33.5430 | 0.8766 | 0.0119 | 38.6587 | 0.9541 |
| Gamma-Reg | 0.0188 | 33.7021 | 0.8802 | 0.0110 | 38.8954 | 0.9579 |
| FBPConvNet | 0.0212 | 33.4847 | 0.8856 | 0.0121 | 37.2277 | 0.9488 |
| LEARN | **0.0113** | **38.9727** | **0.9488** | **0.0082** | **41.7219** | **0.9701** |

Table II. QUANTITATIVE RESULTS ASSOCIATED WITH DIFFERENT ALGORITHMS IN THE THORACIC CASE.

| No. of views | 64 | | | 128 | | |
|---|---|---|---|---|---|---|
| | RMSE | PSNR | SSIM | RMSE | PSNR | SSIM |
| FBP | 0.0696 | 23.1488 | 0.5211 | 0.0403 | 27.8941 | 0.6669 |
| ASD-POCS | 0.0241 | 30.7182 | 0.8857 | 0.0182 | 34.7869 | 0.9395 |
| Dual-DL | 0.0231 | 31.2458 | 0.9020 | 0.0178 | 34.9974 | 0.9451 |
| PWLS-TGV | 0.0198 | 33.5427 | 0.9266 | 0.0108 | 39.2148 | 0.9622 |
| Gamma-Reg | 0.0186 | 33.8451 | 0.9310 | 0.0117 | 38.8987 | 0.9620 |
| FBPConvNet | 0.0190 | 33.3844 | 0.9212 | 0.0127 | 37.8657 | 0.9518 |
| LEARN | **0.0114** | **38.8469** | **0.9650** | **0.0077** | **42.2234** | **0.9806** |

*B. Quantitative and Qualitative Evaluation*

Table I lists the quantitative results for the reconstructions from 64 and 128 views in Fig. 4. It is seen that the proposed LEARN network achieved the best results in terms of the metrics, which agrees with our visual observations. In the cases of 64 and 128 views, our model gained improvements of 5.2 and 3.1 dB for PSNR respectively. Our results were also impressive in terms of RMSE and SSIM.

The quantitative results for the images in Fig. 7 from either 64 or 128 views were presented in Table II. Similar trends can be observed in Table I. The LEARN network preformed the best overall, making consistent improvements in all the metrics.

Table III shows the quantitative results in the full cross validation, obtained by averaging the corresponding values of testing cases. The proposed LEARN network outperformed all the other methods in all the metrics significantly. This solid evidence is in strong agreement with our visual observations. The run time for each method was also given in the table, benchmarked in the CPU mode. Due to the operations invovling the system matrix, the LEARN network was a bit slower than FBPConvNet but LEARN could be viewed as a smarter and faster implementation of iterative algorithms, carries over all the advantages of iterative reconstruction, runs 3-, 300-, 100- and 8-fold faster than ASD-POCS, Dual-DL, PWLS-TGV and Gamma-Reg respectively.

For qualitative evaluation, 30 reference images used for testing and their corresponding under-sampled images reconstructed using different methods were randomly selected for experts' evaluation. Artifact reduction, noise suppression, contrast retention and overall quality were included as qualitative indicators with five assessment grades: from 1 = worst to 5 = best. Two radiologists D1 and D2 respectively with 8 and 6 years of clinical experience scored these images. The reference images were used as the gold standard. For each set of images, the scores were reported as means $\pm$ SDs (average scores $\pm$ standard deviations). The student's t test with $p < 0.05$ was performed to assess the discrepancy. The statistical results are summarized in Table IV.

As demonstrated in Table IV, the impressions on the images reconstructed by FBP were much poorer than that on the reference images in terms of the scores. All the other image reconstruction methods significantly improved the image quality, and PWLS-TGV, Gamma-Reg, FBPConvNet and the proposed LEARN network achieved similar results. The scores of LEARN were closer to the ones of the reference images, and the student's t test results showed a similar trend that the differences between the reference images and the results from LEARN were not statistically significant in all the qualitative indices.

Table III. QUANTITATIVE RESULTS (MEAN) ASSOCIATED WITH DIFFERENT ALGORITHMS IN THE FULL CROSS VALIDATION STUDY.

| No. of views | 64 | | | | 128 | | | |
|---|---|---|---|---|---|---|---|---|
| | RMSE | PSNR | SSIM | Speed | RMSE | PSNR | SSIM | Speed |
| FBP | 0.0684 | 25.1845 | 0.5654 | **0.1047** | 0.0354 | 28.4575 | 0.7358 | **0.1748** |
| ASD-POCS | 0.0215 | 34.1575 | 0.9045 | 15.92 | 0.0172 | 36.1217 | 0.9347 | 23.16 |
| Dual-DL | 0.0221 | 34.2517 | 0.9145 | 1654.25 | 0.0157 | 36.2014 | 0.9425 | 2914.21 |
| PWLS-TGV | 0.0208 | 35.6527 | 0.9286 | 627.24 | 0.0104 | 39.2541 | 0.9607 | 898.45 |
| Gamma-Reg | 0.0202 | 36.0142 | 0.9301 | 50.23 | 0.0120 | 38.7898 | 0.9584 | 86.54 |
| FBPConvNet | 0.0209 | 35.3585 | 0.9267 | 1.08 | 0.0125 | 38.5876 | 0.9533 | 1.93 |
| LEARN | **0.0093** | **40.7337** | **0.9660** | 5.89 | **0.0068** | **43.3812** | **0.9790** | 9.01 |

Table IV. STATISTICAL ANALYSIS OF IMAGE QUALITY SCORES ASSOCIATED WITH DIFFERENT ALGORITHMS (MEAN ±SD).

| | | Reference | FBP | ASD-POCS | Dual-DL | PWLS-TGV | Gamma-Reg | FBPConvNet | LEARN |
|---|---|---|---|---|---|---|---|---|---|
| Artifact reduction | D1 | 3.76±0.43 | 1.23±0.59* | 2.78±0.45* | 2.92±0.56* | 3.30±0.31* | 3.28±0.27* | 3.36±0.25 | **3.52±0.29** |
| | D2 | 3.66±0.33 | 1.15±0.42* | 2.59±0.68* | 2.77±0.87* | 3.26±0.33* | 3.27±0.35* | 3.31±0.35 | **3.49±0.35** |
| Noise suppression | D1 | 3.52±0.68 | 1.64±0.33* | 2.48±0.75* | 2.88±0.52* | 3.27±0.29 | 3.25±0.23 | 3.32±0.18 | **3.45±0.42** |
| | D2 | 3.67±0.55 | 1.45±0.21* | 2.66±0.64* | 2.85±0.66* | 3.25±0.31 | 3.24±0.18 | 3.40±0.20 | **3.52±0.33** |
| Contrast retention | D1 | 3.57±0.41 | 1.27±0.45* | 2.35±0.41* | 2.58±0.66* | 3.22±0.34 | 3.26±0.26 | **3.35±0.26** | 3.30±0.29 |
| | D2 | 3.37±0.67 | 1.17±0.36* | 2.40±0.36* | 2.64±0.47* | 3.10±0.36 | 3.12±0.33 | **3.19±0.20** | 3.15±0.27 |
| Overall image quality | D1 | 3.75±0.41 | 1.12±0.10* | 2.29±0.67* | 2.64±0.54* | 3.25±0.40* | 3.31±0.37 | 3.29±0.38* | **3.50±0.28** |
| | D2 | 3.69±0.36 | 1.09±0.18* | 2.12±0.45* | 2.57±0.67* | 3.17±0.33* | 3.25±0.28 | 3.23±0.35* | **3.46±0.36** |

* indicates $P<0.05$, which means significantly different.

Table V. QUANTITATIVE RESULTS (MEAN) ASSOCIATED WITH DIFFERENT NUMBERS OF FILTERS.

| Num. of Filters | 8 | 16 | 24 | 32 | 48 |
|---|---|---|---|---|---|
| RMSE | 0.0143 | 0.0134 | 0.0115 | 0.0104 | **0.0103** |
| PSNR | 36.9112 | 37.5123 | 38.8135 | 39.7223 | **39.7812** |
| SSIM | 0.9328 | 0.9367 | 0.9514 | 0.9550 | **0.9596** |

Table VI. QUANTITATIVE RESULTS (MEAN) ASSOCIATED WITH DIFFERENT FILTER SIZES.

| Filter Size | 3 | 5 | 7 | 9 |
|---|---|---|---|---|
| RMSE | 0.0115 | **0.0114** | 0.0121 | 0.0124 |
| PSNR | 38.8135 | **38.9543** | 38.4023 | 37.1574 |
| SSIM | 0.9514 | **0.9541** | 0.9507 | 0.9488 |

## C. Trade-Offs between Network and Performance

Although it is believed that one of the advantages of the neural network approach is parameter-free, several parameters of the network architecture are still needed to be set. In practice, with traditional IR methods, such as ASD-POCS, we must adjust the regularization parameter for every task, since the structures of images may be extremely different. On the other hand, with neural networks, the regularization parameters can be learned from training samples, which is highly desirable for image reconstruction targeting a much wider class of tasks. In other words, all the parameters of the LEARN architecture can be selected to match the whole training dataset for multiple tasks. Once the network is trained, it can be applied to all the targeted tasks without further modification.

Specifically, we evaluated the impacts of several key factors of the network, including the number of filters, filter size, number of iterations, the number of training samples and the number of layers in each iteration. The impact of other components of network, including the batch normalization, activation function and loss function were discussed in the supplemental materials. The effect of the parameter was sensed by perturbing one while the others were fixed. The default configuration of the network were $N_t=30$, $n_1=n_2=24$,

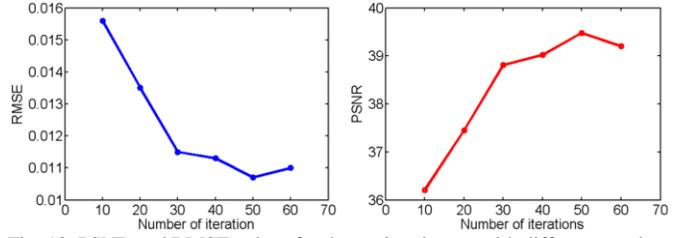

Fig. 10. PSNR and RMSE values for the testing dataset with different numbers of iterations.

$n_3=1$, $s_1=s_2=s_3=3$ and $N_D=100$. In this study, the training and testing sets corresponding to 64 views were chosen and analyzed in the same way as used in the above-described experiments. All the numbers in the following tables or figures denote the average values from all the testing images.

*1) Number of Filters*

We tested the cases of $n_1=n_2=8$, 16, 24, 32 and 48 respectively. The corresponding quantitative results were given in Table V. It can be seen that with the increase of the number of filters, the performance was improved, but the profit gradually declined. Meanwhile, the training and run time will significantly rise. To balance performance and computational time, the number of filters was set to 24 in our LEARN prototype.

*2) Impact of the Filter Size*

Different filter sizes, 3, 5, 7 and 9 respectively, were tested. The results are in Table VI. First, the scores went up with an increased filter size, but when the size was larger than 7, the values of the metrics began to decline. It is well known that increasing the filter size can enlarge the receptive field, which will help CNN extract higher-level features. These features are quite similar to high order statistical features. However, when the filter size increases, more training samples are needed to avoid overfitting. Meanwhile, increasing the filter size will also increase the training and run time.

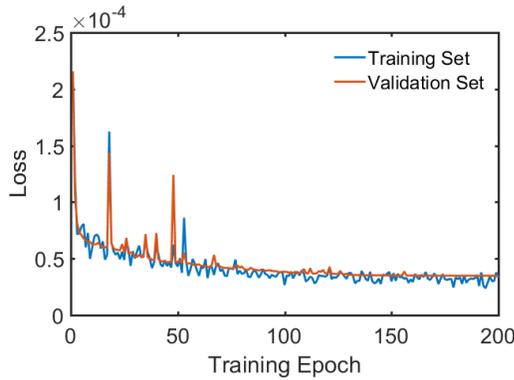

Fig.11. The loss curves of LEARN for training and validation datasets during the training stage.

Table VII. QUANTITATIVE RESULTS (MEAN) ASSOCIATED WITH DIFFERENT NUMBERS OF TRAINING SAMPLES.

| Num. of Samples | 50 | 100 | 150 | 200 | 250 |
|---|---|---|---|---|---|
| RMSE | 0.0147 | 0.0115 | 0.0106 | 0.0106 | **0.0105** |
| PSNR | 36.6812 | 38.8135 | 39.5445 | 39.5637 | **39.6041** |
| SSIM | 0.9309 | 0.9514 | 0.9526 | 0.9574 | **0.9581** |

Table VIII. QUANTITATIVE RESULTS (MEAN) ASSOCIATED WITH DIFFERENT NUMBERS OF LAYERS PER ITERATION.

| Num. of Layers | 2 | 3 | 4 | 5 | 6 |
|---|---|---|---|---|---|
| RMSE | 0.0140 | **0.0115** | **0.0115** | 0.0122 | 0.0121 |
| PSNR | 37.1421 | 38.8135 | **38.8744** | 38.3384 | 38.3746 |
| SSIM | 0.9380 | 0.9514 | **0.9517** | 0.9480 | 0.9470 |

*3) Number of Iterations*

To achieve a satisfactory result, we usually iterate the algorithm a sufficient number of times. Our model can learn the parameters to accelerate the reconstruction procedure, making each LEARN-type iteration more effective than an equivalent iteration in the classic iterative reconstruction process. The quantitative results with different iterations were plotted in Fig. 10. It is seen that when the number of iterations was less than 30, the improvement by adding more iterations was significant. After the number of layers went beyond 30, the performance became saturated. When the number went up to 50, the performance reached the peak and began to decline with additional iterations. Meanwhile, adding more iterations into the network will aggravate the computational burden (more parameters of the network) for training the network. Taking both performance and computational cost into account, we chose $N_t=50$ in our experiments.

*4) Number of Samples*

In traditional applications of deep learning, a huge number of training samples will help improve the performance of the network, but in our work with the utilization of data fidelity in each iteration, the amount of training samples seems not necessary huge. The results with different amounts of training samples are given in Table VII. Clearly, before the number of training samples reached 150, the improvements were significant. After it exceeded 150, the gain diminished greatly. As a result, 200 samples seems adequate for the capacity of our proposed LEARN network. In a general sense, it is with high possibility that over-fitting will be observed while training such a deep network (150 layers) with relatively few samples (200 images in our experiments). Fig. 11 is given to plot the losses with training and validation datasets, and it is suggested in Fig.

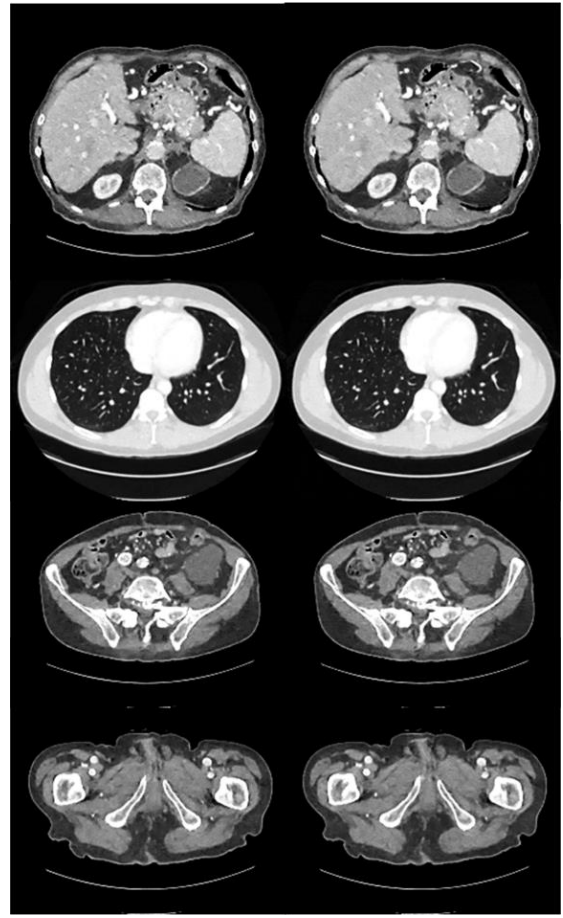

Fig. 12. Four slices reconstructed by LEARN from different initial images. The images in the first column were initialized with the FBP results while the images in the second column were initialized with zero images.

11 that the convergence behavior is fast and stable (the weights were initialized with Gaussian noise of zero mean), and over-fitting is not a problem for our proposed LEARN, which means that 200 examples is a proper number to fit the proposed 150-layer network. In other words, the data was sufficiently large such that over-fitting would not occur, and yet not too huge to be inefficient or impractical. At this moment, there no thorough theory to describe this behavior or find such a sweet spot. Nevertheless, we would like to offer two plausible reasons as follows.

First, the problem we deal with is sparse-view CT reconstruction. The key is to recover local structural details while suppressing artifacts and noise. In contrast to the pattern recognition problems, here we only focus on low-level features, such as edges, shapes, and texture, and 200 images may be already informative to cover these features. There will be no more significant low-level features if more samples are added. Second, different from the traditional CNN architecture, which only has a single input layer, the projection data are directly involved at different layers in LEARN as a strong constraint or regularizer so that large-scale features cannot be much off-target. It can be seen as a simplified version of the densely connected convolutional network (DenseNet) [70], which is proved to have a fast and stable convergence behavior.

Table IX. QUANTITATIVE RESULTS FOR FOUR SLICES AND FULL CROSS VALIDATION OF LEARN WITH DIFFERENT INITIAL IMAGES.

| Initialization | Slice 1 | | Slice 2 | | Slice 3 | | Slice 4 | | Overall | |
|---|---|---|---|---|---|---|---|---|---|---|
| | FBP | Zero | FBP | Zero | FBP | Zero | FBP | Zero | FBP | Zero |
| RMSE | 0.0113 | 0.0129 | 0.0114 | 0.0126 | 0.0091 | 0.0109 | 0.0076 | 0.0094 | 0.0093 | 0.0104 |
| PSNR | 38.9727 | 37.7954 | 38.8469 | 37.9854 | 40.8300 | 39.2764 | 42.3885 | 40.5571 | 40.7337 | 39.3745 |
| SSIM | 0.9488 | 0.9371 | 0.9650 | 0.9592 | 0.9698 | 0.9570 | 0.9750 | 0.9645 | 0.9660 | 0.9588 |

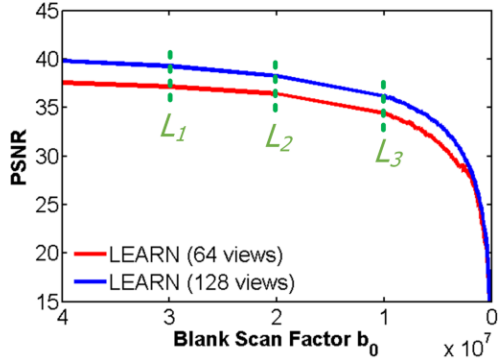

Fig. 13. PSNR values for the reconstructed results using LEARN with different noise levels. The dotted green lines indicate three different noise levels. The corresponding results are in Fig. 13.

*5) Number of Layers for Each Iteration*

To evaluate the impact of the numbers of layers for each iteration, the number of layers per iteration was varied from 2 to 6 with a unit step. The quantitative results are in Table VIII. It can be seen that the performance was significantly improved when the number of layers was greater than 2 and it reached the peak while the number was 4. The performance began to decline after 4. When the number of layers was 6, the number of the network parameters become huge. We had to decrease the batch size to fit the data into the video memory. Based on our feasibility data, 3 layers per iteration seems a decent choice balancing the imaging performance and the computational cost.

*D. Robustness and generalization*

In this subsection, the robustness and generalization of LEARN were also assessed.

*1) Initialization of the iterative process*

Due to the non-convexity of the proposed LEARN network, the initial values may affect the performance. Here we conducted experiments with two typical initialization schemes, which are FBP and zero images respectively, to sense the influence of initialization. Fig. 12 shows four slices from different parts of the human body reconstructed using different initialization schemes. In Fig. 12, it is hard to detect any visual differences between the reconstructed images regardless of the initialization scheme: either FBP or zero images. Table IX lists the quantitative results of these four slices and full cross validations. It is seen that in all the cases, the results with the FBP initial image had a better performance than those with zero initial images, but the differences were not significant, being consistent to the visual inspection. Although the convexity of our model cannot be guaranteed, the merits originated from the proposed framework with learned iteration-wise regularization

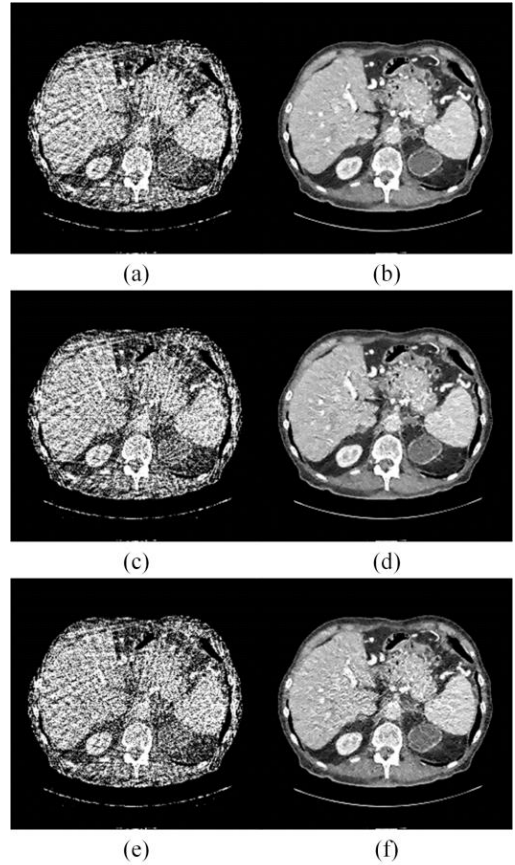

Fig. 14. Reconstructed results from 64 views using FBP and LEARN for different noise levels respectively. (a) FBP result with $b_0 = 3\times10^7$, (b) LEARN result with $b_0 = 3\times10^7$, (c) FBP result with $b_0 = 2\times10^7$, (d) LEARN result with $b_0 = 2\times10^7$, (e) FBP result with $b_0 = 1\times10^7$, and (f) LEARN result with $b_0 = 1\times10^7$.

terms and balancing parameters did yield a competitive performance.

*2) Noise levels*

To show the robustness of the proposed LEARN network against different noise levels, Poisson noise of different strengths was added into the sinograms of testing samples. The networks were trained with noiseless samples from 64 and 128 views respectively. The simulation strategy was similar to that described in [41], and the blank scan factor $b_0$ was set from $4\times10^7$ to $1\times10^5$ with step of $1\times10^5$. Fig. 13 demonstrates the changes of PNSR values associated with different noise levels. It can be seen that the performance of LEARN was quite stable until $b_0$ was decreased below $1\times10^7$. When the noise level was worse than $1\times10^7$, the performance of LEARN began gradually deteriorating. Fig. 14 presents the results of LEARN for three different noise levels, $L_1=3\times10^7$, $L_2=2\times10^7$ and $L_3=1\times10^7$ respectively. In Fig. 14, with the increment of the noise level, the FBP results were degraded heavily. When $b_0$ was greater than $1\times10^7$, the noise was efficiently suppressed using our method, and the image quality of LEARN was judged clinically acceptable. Our results show that the LEARN method can handle a broad range of noise strengths.

Table X. QUANTITATIVE RESULTS (MEAN) ASSOCIATED WITH INCONSISTENCY OF TRAINING AND TESTING SAMPLES.

|  | LEARN-All-T | LEARN-A-T | LEARN-All-A | LEARN-T-A |
|---|---|---|---|---|
| RMSE | 0.0140 | 0.0139 | 0.0100 | 0.0109 |
| PSNR | 37.1999 | 37.1102 | 40.0845 | 39.3153 |
| SSIM | 0.9472 | 0.9484 | 0.9574 | 0.9526 |

*3) Inconsistency of training and testing samples*

It is well known that the performance of learning based methods are usually dependent on training samples. It is meaningful to inspect the robustness relative to training samples. Hence, we trained our network with two different data sources: 200 thoracic and 200 abdominal images. The quantitative results are in Table X. LEARN-All-T and LEARN-All-A denote that the training datasets included different parts of the human body and the corresponding testing datasets contained only thoracic and abdominal images respectively. On the other hand, LEARN-T-A represents the training dataset that only included thoracic images and its corresponding testing dataset only contained abdominal images. LEARN-A-T was just opposite to LEARN-T-A, whose training dataset was composed of only abdominal images and corresponding testing dataset only had thoracic images. Comparing to the results with more diverse training dataset, the results from a single data source had a similar performance. Although some small differences can be noticed, the overall performance was still close to the results with more diverse training dataset. We infer that there would be two reasons for the robustness of our model with respect to training samples: 1) the introduction of projection data helps maintain the reconstruction result within a credible range; and 2) although the structural details of thoracic and abdominal images are quite different, the artifacts caused by the sparse sampling are similar in these cases, and LEARN introduced the residual block that makes the learning procedure of LEARN more focus on artifacts. For these reasons, the ability of the learned filters will not be very sensitive to training samples.

*E. Computational Cost*

Due to the complexity of the proposed network architecture, we implemented it in MatLab aided by MatConvNet. The efficiency of the program can be improved in the current popular frameworks, such as Caffe or TensorFlow. A major issue is how to deal with the huge number of network parameters. For 3 layers in each iteration and 50 iterations in our experiments, the total layers exceed 150 layers and challenges the video memory. For our current implementation, it took 49 hours to train the network with 100 images and 80 hours with 200 images. Although the training stage is time-consuming, the run time is much faster than the classic iterative reconstruction methods, as shown in Table III. Once the training stage is finished offline, the LEARN-based reconstruction is much more efficient than the competing iterative reconstruction methods, be it of the simple TV type or of more advanced, PWLS-TGV, Gamma-Reg, and dictionary learning types.

IV. DISCUSSION AND CONCLUSION

Deep learning has achieved remarkable results in many fields such as computer vision, image analysis, and machine translation. Inspired by this exciting development, it is envisioned that machine learning especially deep learning will play an instrumental role in tomographic reconstruction such as in the field of radiology [52, 71]. Along this direction, learning-based or learned image reconstruction algorithms are being actively developed, focusing on utilizing prior knowledge to improve image quality. Two representative examples are dictionary learning and learned sparsifying transform. In the first example, either a synthesis or an analysis dictionary can be learned from an external data source to represent sparsely an underlying image to be reconstructed. In the second example, a fixed sparsifying transform can be learned from training data. Both examples allow sparse coding data-driven and much more efficient. In this paper, we have demonstrated the feasibility and merits of simultaneously learning multiple transforms/dictionary atoms and associated weighting factors, making the deep learning based image reconstruction more adaptive and more powerful. Essentially, our model can be seen as a generalized version of the previously published learned reconstruction algorithms, with a potential to train the system matrix as well, which not only learns the regularization terms but also all the other parameters in the model. It is underlined that in all the published methods, once the dictionary or the transform was learned, it will not change during the iterative procedure. For the first time, our model learns all regularization terms and parameters in an iteration dependent manner, thereby optimizing image quality and accelerating reconstruction speed in an intelligent way.

Distinguished from most of the previous works, which treated the network structures as black boxes, our model was directly motivated by the numerical scheme for solving the optimization problem based on physical, mathematical and application-relevant knowledge. Hence, the resultant neural network architecture, the LEARN network in this study, was well motivated at the first place, and then optimized via training and testing using the machine learning techniques. This setup offers us unique insights into how the LEARN network achieves its excellent performance. In this sense, we believe that the LEARN network was designed via transfer learning; i.e., some best algorithmic elements from classic iterative reconstruction efforts has been utilized in the LEARN network.

From a general point of view, it is indeed questionable with using a few gradient steps to solve an energy minimization problem, especially for a non-convex energy functional, like (5). Usually, a better way is to consider more advanced optimization algorithms, such as a Quasi-Newton's method or algorithms in [72, 73]. Our original energy functional is truly non-convex since there is a non-convex penalty function in the regularization term. We use a few gradient descent steps for the corresponding energy functional, and this process is finally expressed as a multi-layer CNN model. These gradient descent steps are not to exactly find the global minimum of a specific energy functional. Instead, our intent is to do a few jumps from a starting point, and hopefully we can arrive at a better point closer to the ground truth. As a consequence, the outcome depends on jumping paths, which are controlled by the CNN parameters, as well as the starting point. Under the data fidelity constraints, the training phase is to search such good jumps, and then validated to be successful. In other words, the exploited gradient descent method is used only to derive a very special multi-layer architecture instead of a commonly-used CNN architecture nor a global optimizer. Therefore, it does not matter whether the original energy functional is convex or non-convex, as long as data-driven machine learning yields good results.

In our experiments, our LEARN model has been shown to be generally advantageous in terms of noise suppression, feature preservation, quantitative and qualitative evaluations but image contrast cannot keep perfect in all the situations, such as in Fig. 4. A reason could be due to the utilization of MSE as the loss function, as mentioned in [40]. A perceptual similarity measure may be a better choice for clinical applications. More generally, we can add a discriminative network to measure a more general loss; or in the other words, our current LEARN model can be retrofit into the GAN network, which is currently a hot topic and will be our future effort as well.

In conclusion, motivated by the pioneering results [55, 56] in deep learning for sparse coding [54], we have unfolded a state-of-the-art iterative framework for CT reconstruction into a deep-learning network, called the LEARN network. Except for the system matrix elements, all the other key parameters from the original algorithm have been learned from training samples. We have evaluated the LEARN network with the well-known Mayo Clinic low-dose image dataset in comparison with several state-of-the-art image reconstruction methods. In the experimental results, our LEARN network has demonstrated a favorable performance over the other methods in both image quality and computational efficiency. In our future work, we will further optimize the LEARN network for clinical applications by training the system matrix as well and generalizing the loss function in the GAN framework.

# Learned Experts' Assessment-based Reconstruction Network (LEARN)

This supplement provides additional discussion to accompany our manuscript.

## V. OPTIONAL COMPONENTS FOR THE PROPOSED LEARN NETWORK

It is well known that the network architecture has an important impact on the network performance. Batch normalization (BN), activation and loss functions were altered to assess the impact on the imaging performance. Also, several networks with different architectures were trained using the same strategy as that in Subsection III.C and tested with the same validation dataset. For the activation function, ReLU and parametric ReLU (PReLU) were evaluated. For the loss function, mean squared error (MSE), mean absolute error (MAE), cross entropy (CS) and mean squared logarithmic error (MSLE) were compared. The quantitative results from different configurations of the LEARN network are given in Table XI.

Table XI. THE QUANTITATIVE RESULTS WITH DIFFERENT NETWORK ARCHITECTURES.

| Variant | | RMSE | PSNR | SSIM |
|---|---|---|---|---|
| 1 | LEARN+ReLU+MSE | 0.0115 | 38.8135 | 0.9514 |
| 2 | LEARN+BN+ReLU+MSE | 0.0115 | 38.8304 | 0.9512 |
| 3 | LEARN+PReLU+MSE | 0.0114 | 38.8612 | 0.9533 |
| 4 | LEARN+ReLU+MAE | 0.0123 | 38.2250 | 0.9505 |
| 5 | LEARN+ReLU+MSLE | 0.0122 | 38.3191 | 0.9497 |
| 6 | LEARN+ReLU+CS | 0.0117 | 38.6685 | 0.9517 |

In Table XI, several comments are in order.

i) Comparing Variants. 1 and 2, there was no significant improvement while BN was utilized. The main reason might be that if the expected output of the network satisfies a certain normalized distribution, introducing BN will be beneficial, such as for Gaussian denoising [1]. However, if the expected output obeys a different distribution, BN could produce a compromised outcome. For example, the residual map for image super-resolution is the difference between low-resolution and high-resolution images. Most parts of the residual map are close to zero, and there are some high frequency details across edges and in texture regions. In this situation, BN will decrease the network performance. In [2], the authors removed the BN layers of the network in [3], and the performance was improved. For sparse-view reconstruction, the artifacts and noise in sparse-view CT images have unique characteristics, BN seems not needed as shown by our experimental results.

ii) Comparing Variants 1 and 3, it can be seen that substituting ReLU with PReLU slightly improved the performance. Actually, there is no one activation function that can outperform all other activation functions in all cases. PReLU has a trainable parameter, and improves the model's flexibility. Hence, it is expected that PReLU will be equal or better than ReLU, which is consistent to the experimental results. ReLU was used in our proposed LEARN for the simplicity in this feasibility study.

iii) Comparing Variants 1, 4, 5 and 6, it is found that MSE achieved the best performance among all the four loss functions. It has been shown in many references [33-38, 40-42, 47, 48, 50] that MSE is more suitable for low-level tasks and usually gives better PSNR than other loss functions.

Baes on these observations, a simple network architecture of LEARN (No.1) was used in this manuscript. The reason for this choice is that the purpose of this manuscript is to demonstrate that the merits of the proposed method stemmed from the unfolded IR procedure, learned regularization terms and parameters, not critically depending on a complicated network architecture. A simple network reveals this advantage with little doubt. A more complicated network may improve the performance of LEARN

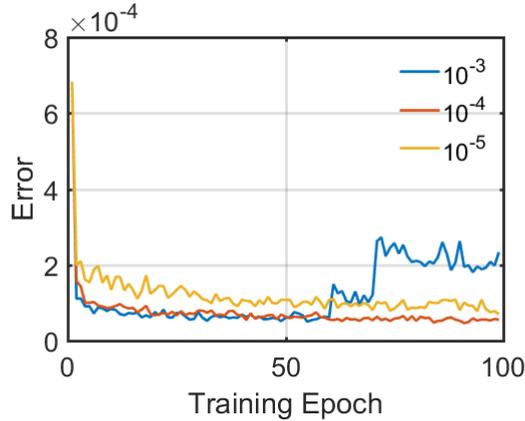

Fig. 15. The loss curves of LEARN for training with different initial learning rates.

further, and it will be a future research topic.

## VI. PARAMETER INITIALIZATION OF THE LEARN NETWORK

The initialization of the network parameters may influence the training and the performance of LEARN. In this study, we manually initialized the parameters, including the learning rate, filter weights and biases. The impacts of different learning rates were shown in Fig. 15. In this study, several networks with different architectures were trained using the same strategy as that used in Subsection III.C and tested with same validation dataset. In Fig. 15, we initialized learning rates of the networks with $10^{-5}$, $10^{-4}$ and $10^{-3}$ respectively. It was seen that when the learning rate was $10^{-3}$, the loss declined more rapidly than the other rates did, but after 60 epochs it began to increase significantly. Theoretically, a greater learning rate may accelerate the training, but there is at the same time a risk of loss increment. In contrast, the curves with smaller learning rates, $10^{-5}$ and $10^{-4}$ converged at a slower speed than the one with the learning rate $10^{-3}$ but the loss was consistently reduced. Meanwhile, the learning rate $10^{-4}$ achieved a better performance than $10^{-5}$. Hence, in our experiments the initial learning rate was set to $10^{-4}$.

Table XII. THE QUANTITATIVE RESULTS WITH DIFFERENT WEIGHT INITIALIZATIONS.

|      | Gaussian | Xavier | MSRA |
|------|----------|--------|------|
| RMSE | 0.0115   | 0.0119 | 0.0117 |
| PSNR | 38.8135  | 38.5599 | 38.7016 |
| SSIM | 0.9514   | 0.9497 | 0.9518 |

In general, the biases are suggested to be initialized to 0. Usually, the weights are initialized randomly according to a certain statistical distribution. In this study, three different weight initialization methods, including Gaussian, Xavier and MSRA were used to evaluate the impact of the weight initialization. The quantitative results with different weight initializations are in Table XII. It can be seen that these results were very close. Although MSRA was originally designed for deep neural networks, it did not help achieve a better performance than the Gaussian initialization. A possible reason is that the proposed LEARN is not a traditional network architecture, it is derived from a numerical scheme for iterative reconstruction, and as strong constraints the projection data were used in every block.